\documentstyle[12pt]{article}

\begin{document}
\large
\title{
A New Algorithm for Numerical Calculation of Link Invariants}
\author{Tetsuo Deguchi and Kyoichi Tsurusaki}
\date{}
\maketitle
\begin{center}
\it
 Department of Physics, Faculty of  Science,\\
     University of Tokyo, Hongo, Bunkyo-ku, Tokyo 113, Japan
\end{center}
\begin{center}
{\bf Abstract}
\end{center}

We introduce an algorithm for numerical calculation of
 derivatives of the Jones polynomial.
This method gives a new  tool for determining
topology of knotted closed loops in three dimensions using
computers.

\newpage
\section{Introduction}

In chemistry and biology, knotted macromolecules such as  DNA knots
can be synthesized in experiments.
\cite{wasserman,sumners}
In statistical mechanics of macromolecules, topological classification
of configurations of polymer chains is an interesting and unsolved problem.
\cite{edwards,saito,wiegel}
Topological constraints could severely restrict available degrees of freedom
in the configuration space,
and would have important effects on  thermodynamic quantities
such as the entropy of the system.
This problem was first  formulated by Delbr{\"u}ck in '60s. \cite{delb}
Since then, several numerical works have been done.
\cite{volo,mehta,michels,koniaris}
In these works unknotting probabilities of closed loops are evaluated
in different systems such as the closed self-avoiding random walks
\cite{mehta}
and the rod-bead model \cite{koniaris}.

In all these numerical researches,
the special values ($\Delta_K(-1)$ or $\Delta_K(2)$)
of the Alexander  polynomial $\Delta_K(t)$
have been used as  tools for classifying knots.
In the algorithms, a closed loop is considered as trivial
if the special value of the Alexander polynomial is found to be one.
However, it is known that the Alexander polynomial does not distinguish
some  knots  from the trivial knot \cite{kinoshita,conway}
(see also \cite{rolfsen}).
Consequently, the special values of the Alexander polynomial are
not complete invariants for classification.

Recently various new link invariants,
such as the Jones, the HOMFLY, and the Kauffman polynomials are introduced.
\cite{jones1,homfly,onknots,pr}
If we could apply these new invariants to computer calculations,
 we would have a  systematic method of
determining topology of closed loops in 3 dimensions.
However, it is not easy to calculate directly any
special value of the new link polynomials
for long closed loops with large number of steps.
If we calculate the link polynomials  through the Markov trace,
 the number of processes in computer calculation  grows
exponentially with respect to the step number $N$ of the loops.
Furthermore, in numerical calculation of link polynomials,
we have to assign  a proper number to the variable $t$ so that
the polynomials do not become numerically too large (overflow).
Thus it seems that we could not directly calculate
the Jones polynomial for large closed loops using computers.

The purpose of this paper is to show
that if we devise new diagrammatic method
using oriented link diagrams,
then derivatives of link polynomials
can be easily calculated using computers.
This approach has the following two advantages.
\begin{enumerate}
\item We can evaluate derivatives of link polynomials
in a calculation time proportional to some power of $N$
(the length of loops).
\item We can avoid the overflow
which could occur in evaluation of polynomials.
\end{enumerate}

\par
The outline of this paper consists of the following.
In \S 2 we introduce a state model for the Jones polynomials
using oriented link diagrams, and  construct
an algorithm for calculating the derivatives.
In \S 3 we show by examples that
the derivatives of the Jones polynomial have enough information
to distinguish knots from the trivial knot.
In \S 4 applying the algorithm we estimate unknotting probability
of closed self-avoiding random walks.

\section{Algorithm of calculating  the derivatives }
\subsection{Oriented state model}
Let us  consider diagrammatic methods for calculation of  link polynomials.
We call the methods state models.
We have  two types of state models of the Jones polynomial,
that of unoriented link diagrams, and that of oriented link diagrams.
We call a state model using oriented (unoriented) link diagrams
oriented (unoriented) state model.
L.H. Kauffman introduced an unoriented diagrammatic method
called bracket polynomial
for the Jones polynomial. \cite{statemodel}
It seems, however, that
the bracket polynomial is not suitable
for calculation of derivatives of the Jones polynomial using computers.

We formulate  an oriented state model of the Jones polynomial.
\cite{turaev,ado}
Any oriented link diagram can be decomposed into the
tangle diagrams given in Fig. 2.1.
\begin{center}
Fig. 2.1
\end{center}
For the tangle diagrams we define diagram weights.
The nonzero diagram weights are given by the following.
\begin{eqnarray}
 U_{1 2} & = & t^2 , U_{2 1}=1, U^{1 2}= 1 , U^{2 1}=t^{-2} ,
\label{eqn:U} \\
G(+;1)^{1 1}_{1 1} \!\!\! &=& \!\!\! G(+;1)^{2 2}_{2 2}=1 ,
G(+;1)^{1 2}_{2 1}=G(+;1)^{2 1}_{1 2}=t^2 ,
\nonumber \\
G(+;1)^{1 2}_{1 2}\!\!\! &= & \!\!\! 1-t^4 ,
G(+;2)^{1 1}_{1 1} = G(+;2)^{2 2}_{2 2}=t^2 ,
\nonumber \\
G(+;2)^{1 2}_{2 1}\!\!\!& = & \!\!\!  G(+;2)^{2 1}_{1 2}=1 ,
G(+;2)^{2 1}_{2 1}  =   t^{-2}-t^2 ,
\nonumber \\
G(-;1)^{1 1}_{1 1} \!\!\!& = & \!\!\! G(-;1)^{2 2}_{2 2}=1 ,
G(-;1)^{1 2}_{2 1}  = G(-;1)^{2 1}_{1 2}=t^{-2} ,
\nonumber \\
G(-;1)^{2 1}_{2 1} \!\!\! &= & \!\!\! 1-t^{-4} ,
G(-;2)^{1 1}_{1 1} =G(-;2)^{2 2}_{2 2}=t^{-2} ,
\nonumber \\
G(-;2)^{1 2}_{2 1}\!\!\! & = & \!\!\! G(-;2)^{2 1}_{1 2}=1 ,
G(-;2)^{1 2}_{1 2} = t^2-t^{-2} .
 \label{eqn:G}
\end{eqnarray}
The other nonzero diagram weights are given by the following relations.
\begin{eqnarray}
 U_{a b}&=& \widetilde{U}_{a b} , \quad U^{a b}=\widetilde{U}^{a b},
 \nonumber \\
G(\pm;1)& =& G(\pm;3) , \quad G(\pm;2)=G(\pm;4) .
\end{eqnarray}

We introduce matrix notation  $X=X^{ab}_{cd}E^a_c \otimes E^b_d$, where
$(E^a_c)_{ij} = \delta_{ai} \delta_{cj}$.  For example we have
\begin{equation}
 U_{a b}=\widetilde{U}_{a b}=\left(
\begin{array}{cc}
0 & t^2 \\
1 & 0
\end{array}
\right), \\
\end{equation}
\begin{equation}
G(+;1)=G(+;3) =
\left( \begin{array} {cccc}
1 & 0 & 0 & 0 \\
0 & 1-t^4 & t^2 & 0 \\
0 & t^2 & 0 & 0 \\
0 & 0 & 0 & 1
\end{array} \right). \\
\end{equation}

We now consider state sum of the oriented state model.
We can represent any given link $L$  by  a link diagram.
We  decompose  the link diagram
into pieces of the oriented tangle diagrams given in Fig. 2.1.
We put variables on the edges of the tangle diagrams  (see Fig. 2.2).
\begin{center}
Fig. 2.2
\end{center}
The variables can take two values, 1 and 2.
Let each one of the variables take either 1 or 2,  then
 we have a configuration (of variables) on the link diagram.
We now assign the diagram weights in eqs. (\ref{eqn:U}) and
(\ref{eqn:G}) to the oriented tangle diagrams
 and take multiplication of the diagram weights.
Then we take summation over all the possible configurations
on the link diagram. Let the symbol $\phi(L)$ denote the sum.
We call the sum $\phi(L)$ oriented state sum. For example, we give
the oriented state sum for the knot $3_1$ (see Fig. 2.2).
\begin{eqnarray}
\label{eqn:bracket}
\phi(L)=G^{a b}_{e f}(-;1) \cdot G^{c d}_{g h}(-;3) \cdot G^{f g}_{i j}(-;4)
\cdot U_{a d} \cdot U_{a c} \cdot \widetilde{U}^{e i} \cdot \widetilde{U}^{j h}
{}.
\end{eqnarray}
Here we have assumed  the Einstein notation for sums over repeated indices.
Another quantity necessary for calculation of the link polynomial
is the writhing number. \cite{onknots}
The writhing number is defined as follows.
The oriented crossings have two types ($G(+)$ and $G(-)$).
Each of these crossing types has been labeled with +1 or -1.
The writhe of a link is defined by the sum of the labels for all the crossings
in the link diagram. We thus define the Jones  polynomial $\alpha_J(L)$
using $\phi(L)$ and $writhe$ by
\begin{equation}
\alpha_J(L)=\frac{\phi(L)}{t^2+t^{-2}} \cdot (t^2)^{writhe} .
\label{eqn:def}
\end{equation}
By checking the Reidemeister moves of the tangle diagrams \cite{turaev}
we can show that the definition (\ref{eqn:def}) gives an isotopy
invariant of knots and links. It is the Jones polynomial.

The oriented state model has the advantage that all the nonzero entries
of the diagram weights of $U$ and $\widetilde{U}$
do not involve the $\pm$ sign although the nonzero entries are off-diagonal.
If we take the limit $t \rightarrow 1$,
then the oriented state model satisfies the following properties.
\begin{eqnarray}
\label{eqn:proper}
\lim_{t \rightarrow 1}\!\!\!\!\!\!&U&\!\!\!\!\!\!
= \lim_{t \rightarrow 1}\widetilde{U}
=\!\!\!
\left(
\begin{array} {cc}
0 & 1 \\
1 & 0
\end{array}
\right),  \nonumber \\
& & \nonumber \\
\lim_{t \rightarrow 1}\!\!\!&G^{ab}_{cd}&\!\!\!(k;\pm)
= \delta^a_d \delta^b_c \hspace{1cm}  (\mbox{for } \hspace{2mm} k=1,2,3,4) .
\end{eqnarray}
where $\delta^a_d$ and $\delta^b_c$ denote the Kronecker delta.

Let us compare the oriented state
model with the unoriented one.
If we use the unoriented state model
(bracket polynomial),
 in the limit $t \rightarrow 1$,
the nonzero entries  of the matrices $U$ have  $\pm$ factors.
Therefore  we have to count how many times the signs $\pm$ occur
in a given configuration, and thus calculation time will grow
as fast as $2^N$ with respect to the system size $N$.

\subsection{Expansion of the state sum}
Let us discuss calculation of derivatives  of the Jones polynomial.
We set
\begin{equation}
t^2 = 1+\varepsilon,
\end{equation}
and consider expansion of the Jones polynomial
with respect to $\varepsilon$.
Because the Jones polynomial is link invariant,
the  coefficients are also link invariants.
The oriented state model in the last section directly leads to
an algorithm for calculation of the coefficients.
We expand the matrices $G$ and $U$ in terms of the parameter
$\varepsilon$.  We calculate the oriented state sum using the expanded
matrices. For an illustration
we consider the matrices $G(+;1)$ and $U_{a b}$
\begin{eqnarray}
 G^{a b}_{c d}(+;1)\!\!\!&=&\!\!\!\delta^a_d \delta^b_c
+\varepsilon
\cdot G^{a b}_{c d}(+;1)^{(1)}
+\varepsilon^2
\cdot G^{a b}_{c d}(+;1)^{(2)} \nonumber \\
\!\!\!&=&\!\!\!
\left( \begin{array} {cccc}
1 & 0 & 0 & 0 \\
0 & 0 & 1 & 0 \\
0 & 1 & 0 & 0 \\
0 & 0 & 0 & 1
\end{array} \right)+\varepsilon
\left( \begin{array} {cccc}
0 & 0 & 0 & 0 \\
0 & -2 & 1 & 0 \\
0 & 1 & 0 & 0 \\
0 & 0 & 0 & 0
\end{array} \right)
+ \varepsilon^2
\left( \begin{array} {cccc}
0 & 0 & 0 & 0 \\
0 &-1 & 0 & 0 \\
0 & 0 & 0 & 0 \\
0 & 0 & 0 & 0
\end{array} \right), \nonumber \\
U_{a b}\!\!\!&=&\!\!\!U^{(0)}_{a b}+\varepsilon\cdot U^{(1)}_{a b}
\hspace{7.5cm} \nonumber \\
\!\!\!&=&\!\!\!
\left( \begin{array} {cc}
0 & 1 \\
1 & 0
\end{array} \right)
+\varepsilon
\left( \begin{array} {cc}
0 & 1 \\
0 & 0
\end{array} \right) .
\end{eqnarray}
We call the matrices  of 0-th, 1-st, 2-nd orders
$\cdots$ in the $\varepsilon$ expansion
matrices of 0-th order, 1-st order, 2-nd order, $\cdots$, respectively.

Let us show how to evaluate  the second order contributions
in the oriented state sum for the knot $3_1$ (Fig 2.2).
We first expand $\phi(L)$ in eq. (\ref{eqn:bracket}) by $\varepsilon$.
\begin{eqnarray}
\phi(L)\!\!\!&=&\!\!\!2 \nonumber \\
\!\!\!&+&\!\!\!\varepsilon \biggl\{
G^{a b}_{e f}(-;1)^{(1)} \delta^c_h \delta^d_g  \delta^f_j \delta^g_i
U^{(0)}_{a d}  U^{(0)}_{b c}  \widetilde{U}^{e i(0)}
\widetilde{U}^{j h(0)}+\cdots \biggr\} \nonumber \\
\!\!\!&+&\!\!\!\varepsilon^2 \biggl\{
G^{a b}_{e f}(-;1)^{(1)}  G^{c d}_{g h}(-;3)^{(1)}  \delta^f_j\delta^g_i
U^{(0)}_{a d}  U^{(0)}_{b c}  \widetilde{U}^{e i(0)}
\widetilde{U}^{j h(0)} \nonumber \\
\!\!\!&+&\!\!\!
G^{a b}_{e f}(-;1)^{(1)}  \delta^c_h \delta^d_g
G^{f d}_{i j}(-;4)^{(1)}  U^{(0)}_{a d}  U^{(0)}_{b c}
\widetilde{U}^{e i(0)}  \widetilde{U}^{j h(0)} + \cdots \biggr\}
\nonumber \\
\qquad &+& \!\!o(\varepsilon^2) .
\label{eqn:expand}
\end{eqnarray}
We now consider the first term of the 2-nd order
$G^{a b}_{e f}(-;1)^{(1)} \cdot$
\par \noindent
$G^{c d}_{g h}(-;3)^{(1)} \cdots$  in the expansion (\ref{eqn:expand}).
This term is equivalent to the  configuration of Fig 2.3
where the two  matrices enclosed with  circles
 are 1-st order  and all the other are 0-th order.

We can readily evaluate the term
without directly calculating  the oriented state sum
with respect to  all the variables on the diagram.
We identify the two variables on those edges of the braiding diagrams
that are connected by the dotted lines each other,
and then  take the sum for the remaining independent variables.
For any link diagram  we can thus reduce
the number of independent variables into at most 4
in calculation  of the second order contributions.
Thus we have the following
\begin{eqnarray}
  \sum_{a,b,c,d,e,f,g,h} \!\!\!\!& &\!\!\!\!
G^{a b}_{e f}(-;1)^{(1)} \cdot G^{c d}_{g h}(-;3)^{(1)}
\cdot \delta^f_j \delta^g_i \cdot U^{(0)}_{a d} \cdot U^{(0)}_{b c} \cdot
\widetilde{U}^{e i(0)} \cdot \widetilde{U}^{j h(0)} \nonumber \\
= \sum_{a,b,g,h}{}^{'} \!\!\!\!& &\!\!\!\! G^{a b}_{e f}(-;1)^{(1)} \cdot G^{c
d}_{g h}(-;3)^{(1)}
=-4 . \label{eqn:delete}
\end{eqnarray}
Here the symbol $\sum_{a,b,g,h}^{'}$ means that we take the summation
for the independent variables
$a,b,g,h$ where $c,d,e$ and $f$ are assumed to be
conjugate of $b,a,g$ and $h$, respectively; when $a=1$, then $d=2$.
It is important to note that in calculation
of the term in (\ref{eqn:delete})
we can eliminate the parts of  the Kronecker deltas
and the matrices $U^{(0)}$
because of the property (\ref{eqn:proper}).

To finish this section,   we make a comment.
We can apply the method in this section to other link polynomials.
We can calculate derivatives of the HOMFLY polynomial,
in particular, of the Alexander polynomial.
For an illustration, the oriented state model
for the Alexander polynomial is given in Appendix A.

\section{On derivatives of the Jones polynomial}
\subsection{Tools of classifying knots }

We give a table of the derivatives of the Jones polynomial
and the special value of the Alexander polynomial ($\Delta_K(-1)$) for knots.
\begin{center}
Table 3.1
\end{center}
It is easy to see from the table that derivatives of the Jones polynomial are
useful in classifying  knot types.
Since link polynomials are Laurent polynomials of a variable $t$,
we can recover link polynomials from their derivatives (if we
previously know something about the links such
as  upper bound of the crossing number etc.).

You may think that the second derivative of the Jones polynomial
is not sufficient to distinguish a non-trivial knot from different knots.
However, if we calculate the third order,
then we may have much more precise distinction of knots
(at least as far as prime knots are concerned).
Further, if we calculate the fourth order we can distinguish
even such knots that cannot
be distinguished by the Alexander polynomial.

Let us consider a composite knot $(3_1)(3_1)(3_1)$ and
a prime knot $8_{10}$ that share the same Alexander polynomial
(see e.g., Ref. \cite{koniaris}).
{}From
Table 3.1 we see that
the fourth derivative of the Jones polynomial distinguishes
the two knots while
for the Alexander polynomial at $t=-1$
($\Delta_K(-1)$)  the values for knots $(3_1)(3_1)(3_1)$ and $8_{10}$ coincide.
When we calculate
the Alexander and the Jones polynomials for  composite knots (links)
it is useful that the link polynomial for the composite link
$L_1 \# L_2$ is given by $\alpha(L_1 \# L_2)
= \alpha(L_1)\alpha(L_2)$.

We think that the algorithm can make up for possible
faults of the Alexander polynomial
and thus we can determine knot types of closed loops
much more exactly.

\subsection{Remarks on derivatives of the Jones polynomial}

It is noteworthy that
derivatives of the Jones polynomial have many applications in knot theory.
The following formulas have been given. \cite{hmurakami}
\begin{eqnarray}
\label{1order}
{\frac {d} {dt}} V_K(t=1)\!\!\!&=&\!\!\!0,  \\
& & \nonumber \\
{\frac {d^2} {dt^2}} V_K(t=1)\!\!\!&=&\!\!\! \mbox{const.}\times a_2(K) .
\end{eqnarray}
Here $a_2(K)$ is the second coefficient of
the Conway polynomial (the Alexander polynomial) \cite{conway},
which is related to {\it Arf invariant}.  \cite{onknots,hmurakami}

We may discuss  derivatives of the Jones polynomial
from the viewpoint of perturbation theory of the Chern-Simons field theory.
 The coefficients for the terms
in the perturbational expansion of the Wilson lines are expressed
in terms of the Feynman integrals. \cite{martellini}
The numerical calculation of derivatives of the Jones polynomial corresponds
to evaluation of the Feynman integrals.

\section{Numerical results and concluding remarks}

We can apply the algorithm introduced in the present paper
to any system of closed loops.
For an illustration using the second derivative of the Jones polynomial
 we analyzed  the system of rings \cite{mehta}
 generated  by closed Gaussian random walks.
\begin{center}
Graph 4.1
\end{center}
The probability of occurrence of the trivial knot is
plotted in the logarithmic scale
as a function of the number of steps ($N$) of generated closed loops.
We call the probability of the trivial knot unknotting
probability.
With the number of steps $N$  fixed we evaluate  the unknotting probability
$P_0=N_t/M$ by counting the number $N_t$ of trivial knots
when we generate $M$= 1000 closed random walks.

{}From Graph 4.1  we see that the unknotting probability has  an
exponentially decaying behavior $P_0(N) \sim \exp(-N/N_0)$, where
we call  $N_0$ characteristic length.
We may consider  $N_0$  necessary steps to form a nontrivial knot.
{}From Graph 4.1.  we  have an estimation of  characteristic length $N_0$;
$300 < N_0 < 370$.

Let us consider efficiency of our algorithm.
We plot the time necessary for calculation of the second coefficients
as a function of the number of steps ($N$) of given closed loops.
\begin{center}
Graph 4.2
\end{center}
We see from Graph 4.2 that
the calculation time for the second derivative of the Jones polynomial
behaves asymptotically as  $N^2$ with respect to  $N$.
{}From the result we may expect that
the time for calculation of the $r$-th derivative grows as $N^r$, or
at least non-exponentially with respect to $N$.

This result makes a clear difference from the other approaches
to the Jones polynomial.  We can calculate
the Jones polynomial through explicit evaluation of
the Markov trace or diagrammatic calculation  using the skein relation.
If these methods are applied to computer calculations,
then they yield exponentially growing calculation time with respect to $N$.
We thus conclude that derivatives of link polynomials
give a simple and systematic computer-orientated method
of determining topology of given closed loops in 3 dimensions
and that the algorithm in the paper can be applied to various problems
related to knotted configurations in statistical physics
and many body problem.

{\vskip 2.0cm}

\par \noindent
{\bf Acknowledgements}
\par
We would like to thank Prof. M. Wadati for his keen interest in this work.

\newpage
\setcounter{equation}{0}
\renewcommand{\theequation}{A.\arabic{equation}}
\appendix
\section{Appendix}
We introduce an oriented state model of
the Alexander polynomial.  \cite{deg,ado}
The diagram weights $U_{ab} \cdots$,
 and $G(\pm;k)^{ab}_{cd}$ are defined
for the tangle diagrams in Fig. 2.1.
The nonzero diagram weights are given by the following.
\begin{equation}
U_{1 1} = U^{1 1}=\widetilde{U}_{1 1}=\widetilde{U}^{1 1}=\widetilde{U}_{2 2}
=\widetilde{U}^{2 2}=1 , U_{2 2}=U^{2 2}=-1 ,
\end{equation}
\begin{eqnarray}
G(+;1)^{1 1}_{1 1} \!\!&=&\!\! 1,
G(+;1)^{1 2}_{2 1}=G(+;1)^{2 1}_{1 2}=t^{-2},
G(+;1)^{2 1}_{2 1}=1-t^{-4}, \nonumber \\
G(+;1)^{2 2}_{2 2} \!\!&=&\!\! -t^{-4},G(+;2)^{1 1}_{1 1}=1,G(+;2)^{1 2}_{2
1}=G(+;2)^{2 1}_{1 2}=t^{-2}, \nonumber \\
G(+;2)^{2 2}_{1 1} \!\!&=&\!\! t^{-4}-1,G(+;2)^{2 2}_{2 2}=-t^{-4},G(+;3)^{1
1}_{1 1}=1, \nonumber \\
G(+;3)^{1 2}_{2 1} \!\!&=&\!\! G(+;3)^{2 1}_{1 2}=t^{-2},G(+;3)^{1 2}_{1
2}=1-t^{-4}, \nonumber \\
G(+;3)^{2 2}_{2 2} \!\!&=&\!\! -t^{-4},G(+;4)^{1 1}_{1 1}=1,G(+;4)^{1 2}_{2
1}=G(+;4)^{2 1}_{1 2}=t^{-2}, \nonumber \\
G(+;4)^{1 1}_{2 2} \!\!&=&\!\! 1-t^{-4},G(+;4)^{2 2}_{2 2}= -t^{-4},G(-;1)^{1
1}_{1 1}=1, \nonumber \\
G(-;1)^{1 2}_{2 1} \!\!&=&\!\! G(-;1)^{2 1}_{1 2}=t^2,G(-;1)^{1 2}_{1 2}=1-t^4,
\nonumber \\
G(-;1)^{2 2}_{2 2} \!\!&=&\!\! -t^4,G(-;2)^{1 1}_{1 1}=1,G(-;2)^{1 2}_{2
1}=G(-;2)^{2 1}_{1 2}=t^2, \nonumber \\
G(-;2)^{1 1}_{2 2} \!\!&=&\!\! t^4-1,G(-;2)^{2 2}_{2 2}=-t^4, G(-;3)^{1 1}_{1
1}=1, \nonumber \\
G(-;3)^{1 2}_{2 1} \!\!&=&\!\! G(-;3)^{2 1}_{1 2}=t^2,G(-;3)^{2 1}_{2
1}=1-t^4,G(-;3)^{2 2}_{2 2}=-t^4,\nonumber \\
 G(-;4)^{1 1}_{1 1} \!\!&=&\!\! 1,G(-;4)^{1 2}_{2 1}=G(-;4)^{2 1}_{1
2}=t^2,G(-;4)^{2 2}_{1 1}=1-t^4, \nonumber \\
G(-;4)^{2 2}_{2 2} \!\!&=&\!\! -t^4.
\end{eqnarray}

By the oriented state model we
can calculate derivatives of the Alexander polynomial. We give two remarks.
(1) $U^{ab}\cdots $, are diagonal matrices, although there are $-1$ in
some of the nonzero entries.  (2) The nonzero off-diagonal elements of the
matrices $G(\pm:j)$  become 1  when $t \rightarrow 1$
(cf. eq. (\ref{eqn:proper})).

\newpage
\noindent
{\bf Figure Captions}

\par \noindent
Fig. 2.1  Oriented tangle diagrams.

\par \noindent
Fig. 2.2  Link diagram of knot $3_1$ and variables $a,b,c,d,e,f,g,h,i,j$.

\par \noindent
Fig. 2.3  Configuration for the term
\par \[ \sum_{a,b,c,d,e,f,g,h}
G^{a b}_{e f}(+;1)^{(1)} \cdot G^{c d}_{g h}(+;3)^{(1)}
\cdot \delta^f_j \delta^g_i \cdot U^{(0)}_{a d} \cdot U^{(0)}_{b c} \cdot
\widetilde{U}^{e i(0)} \cdot \widetilde{U}^{j h(0)} \]

\par \noindent
Table 3.1 The symbol $\Delta_K(-1)$ denotes the Alexander polynomial.
Remark that the first derivative of the Jones polynomial for a knot
vanishes (see the formula (\ref{1order})).

\par \noindent
Graph 4.1 Unknotting probability $P_0(N)$ as a function of length $N$.

\par \noindent
Graph 4.2 Calculation time versus length $N$.

\end{document}